\documentclass[twocolumn]{revtex4}

\usepackage{amsfonts,amssymb,amsbsy,bm,graphicx}

\begin{document}

\title{General unit-disk representation for
periodic multilayers}

\author{Alberto G. Barriuso, Juan J. Monz\'on,
and Luis L. S\'anchez-Soto}

\affiliation{Departamento de \'Optica,
Facultad de F\'{\i}sica,
Universidad Complutense,
28040 Madrid, Spain}

\begin{abstract}
We suggest a geometrical framework to discuss
periodic layered structures in the unit disk.
The band gaps appear when the point representing
the system approaches the unit circle. We show
that the trace of the matrix describing the basic
period allows for a classification in three families
of orbits with quite different properties. The laws
of convergence of the iterates to the unit circle
can be then considered as universal features of
the reflection.
\end{abstract}


\maketitle

\newpage

Photonic band gap structures~\cite{dowling}
can be dealt with by computing the band
structure (using e.g. Bloch theory) or from
the point of view of scattering~\cite{Cen00}.
The essential difference is that a scattering
experiment always involves a finite structure,
while Bloch waves imply an endlessly repetition
of the basic period.

In the context of electromagnetic optics,
photonic crystals (that is, one-dimensional
periodic layered structures) have attracted
recently a lot of attention because the striking
property of acting as omnidirectional reflectors:
they can reflect light at any polarization,
any incidence angle, and over a wide range of
wavelengths~\cite{Fin98,Dow98,Yab98,Chi99,Lek00}.

The appearance of strong reflection (stop bands)
depends on the properties of the basic period.
Different theoretical approaches, involving
equivalent medium theories, group velocity,
and other quantities~\cite{Gra00,Mot00}, have
been developed to account for the detailed
structure of these stop bands and their edges.
Each one of these models emphasizes some
aspects of the problem but, at the same
time, has some drawbacks.

In the present Letter we introduce a
geometrical setting that allows for a
deeper understanding of periodic systems.
Our treatment is quite general and only
assumes linearity: it applies to any physical
system whose transfer matrix belongs to the
group SU(1,~1). The key point for our purposes
is the fact that the multilayer transfer
function induces a bilinear transformation
in the unit disk~\cite{Yon02,Mon02}. Since
perfect mirrors are represented precisely
by the unit circle, the route to a stop
band can be understood as the convergence
of the point representing the action of
the system to the unit circle.

We start by examining the basic period of
the system, which consists of a stack of
plane-parallel layers sandwiched between
two semi-infinite ambient ($a$) and substrate
($s$) media that we shall assume to be
identical, since this is the common
experimental case. Hereafter all the
media are supposed to be lossless,
homogeneous, and isotropic.

A monochromatic linearly polarized plane
wave falls from the ambient making an angle
$\theta_0$ with the normal to the first
interface and with an amplitude $E_{a}^{(+)}$.
We consider as well another plane wave of the
same frequency and polarization, and with amplitude
$E_{s}^{(-)}$, incident from the substrate at
the same angle $\theta_0$. The output fields in
the ambient and the substrate will be denoted
$E_{a}^{(-)}$ and $E_{s}^{(+)}$, respectively.

The field amplitudes at each side of the
multilayer are related by the linear relation
\begin{equation}
\label{Evec}
\left (
\begin{array}{c}
E_a^{(+)} \\
E_a^{(-)}
\end{array}
\right )
= \mathsf{M}_{as}
\left (
\begin{array}{c}
E_s^{(+)} \\
E_s^{(-)}
\end{array}
\right ) ,
\end{equation}
where the multilayer transfer matrix $\mathsf{M}_{as}$
can be shown to be~\cite{Mon99a,Mon99b}
\begin{equation}
\label{Mlossless}
\mathsf{M}_{as} = \left [
\begin{array}{cc}
1/T_{as} & R _{as}^\ast/T_{as}^\ast \\
R_{as}/T_{as} & 1/T_{as}^\ast
\end{array}
\right ] \equiv \left [
\begin{array}{cc}
\alpha & \beta \\
\beta^\ast & \alpha^\ast
\end{array}
\right ] .
\end{equation}
Here the complex numbers $R_{as}$ and $T_{as}$
are, respectively, the overall reflection and
transmission coefficients for a wave incident
from the ambient. Because $ |R _{as}|^2 +
|T _{as}|^2 =1$, we have $\det \mathsf{M}_{as}
= +1$ and then the set of lossless multilayer
matrices reduces to the group SU(1,~1).

We are often interested in the transformation
properties of field quotients rather than
the fields themselves.  Therefore, we
introduce the complex numbers
\begin{equation}
\label{defz}
z  = \frac {E^{(-)}}{E^{(+)}} ,
\end{equation}
for both ambient and substrate. Equation~(\ref{Evec})
defines then a transformation on the complex
plane ${\mathbb{C}}$, mapping the point
$z_s$ into the point $z_a$, according to
\begin{equation}
\label{accion}
z_a = \Phi [\mathsf{M}_{as} , z_s] =
\frac{\beta^\ast +\alpha^\ast z_s}
{\alpha + \beta z_s} ,
\end{equation}
which is a bilinear (or M\"{o}bius) transformation.
This action can be seen as a function $z_a = f(z_s)$
that will be called the multilayer transfer
function. One can check that the unit disk,
the external region and the unit circle remain
invariant under the multilayer action~\cite{Yon02}.
Note that $|z_a| = |z_s| = 1$ for a perfect
mirror with light incident from both the
ambient and the substrate, so the transformation
(\ref{accion}) relates then points on the unit
circle. When no light strikes from the substrate
$z_s = 0$ and $|z_a| = 1$, so a mirror maps
the origin into a point on the unit circle.

In what follows, the idea of fixed points of the
transformation will prove to be essential. These
points can be defined as the field configurations
such that $z_a = z_s \equiv z_f$ in Eq.~(\ref{accion});
i.e., $z_f = \Phi [\mathsf{M}_{as}, z_f]$, whose
solutions are
\begin{equation}
z_{f \pm} = \frac{1}{2 \beta}
\left \{  -2 i \ \mathrm{Im}(\alpha) \pm
\sqrt{[\mathrm{Tr} ( \mathsf{M}_{as} )]^2 -4} \right \} .
\end{equation}
The trace of $\mathsf{M}_{as}$ provides then
a suitable tool for the classification of
multilayers~\cite{San01}.

When $ [\mathrm{Tr} ( \mathsf{M}_{as} )] ^2 < 4$
the multilayer action is elliptic and it has only
one fixed point inside the unit disk, while the
other lies outside. When $ [ \mathrm{Tr} (
\mathsf{M}_{as} )]^2 > 4$ the action is
hyperbolic and it has two fixed points on
the unit circle. Finally, when $ [ \mathrm{Tr}
(\mathsf{M}_{as}) ]^2 = 4$ the multilayer action
is  parabolic and it has only one (double)
fixed point on the unit circle.

To proceed further let us note that by taking the
conjugate of $\mathsf{M}_{as}$ with any matrix
$\mathsf{C}\in $ SU(1,~1), that is
\begin{equation}
\label{conjC}
\widehat{\mathsf{M}}_{as} = \mathsf{C} \
\mathsf{M}_{as} \ {\mathsf{C}}^{-1} ,
\end{equation}
we obtain another matrix of the same type,
since $\mathrm{Tr} (\widehat{\mathsf{M}}_{as}) =
\mathrm{Tr} (\mathsf{M}_{as})$. Conversely,
if two multilayer matrices have the same trace,
a matrix $\mathsf{C}$ satisfying Eq.~(\ref{conjC})
can be always found.

The fixed points of $\widehat{\mathsf{M}}_{as}$
are then the image by $\mathsf{C}$ of the fixed
points of $\mathsf{M}_{as}$. In consequence, given
any matrix $\mathsf{M}_{as}$, it can always be
reduced to a unique $\widehat{\mathsf{M}}_{as}$
with one of the following canonical forms:
\begin{eqnarray}
\label{Iwasa1}
\widehat{\mathsf{K}} (\varphi) & = & \left [
\begin{array}{cc}
\exp (i\varphi/2) & 0 \\
0 & \exp (-i\varphi/2)
\end{array}
\right ]\ ,
\nonumber \\
\widehat{\mathsf{A}}(\chi) & = & \left [
\begin{array}{cc}
\cosh (\chi/2) & i\, \sinh(\chi/2) \\
-i\, \sinh(\chi/2) & \cosh (\chi/2)
\end{array}
\right ]\ , \\
\widehat{\mathsf{N}} (\eta) & = &
 \left [
\begin{array}{cc}
1 - i \eta/2 & \eta/2 \\
\eta/2 & 1+ i \eta/2
\end{array}
\right ]\ , \nonumber
\end{eqnarray}
that have as fixed points the origin (elliptic),
$+i$ and $-i$ (hyperbolic) and $+i$ (parabolic),
and whose physical significance has been
studied before~\cite{Mon01}.

\begin{figure}
\centering
\resizebox{0.75\columnwidth}{!}{\includegraphics{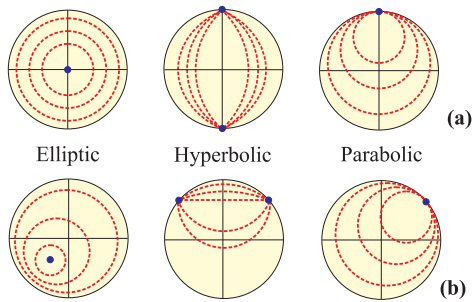}}
\caption{Plot of typical orbits in the unit disk for: (a)
canonical transfer matrices as given in Eq.~(\ref{Iwasa1}) and (b)
arbitrary transfer matrices.}
\end{figure}

The concept of orbit is especially appropriate for
obtaining a picture of these actions. Given a point
$z$, its orbit is the set of points $z^\prime$
obtained from $z$ by the action of all the
elements of the group. In Fig.~1.a we have plotted
typical orbits for each one of the canonical forms.
For matrices $\widehat{\mathsf{K}}(\varphi)$ the
orbits are circumferences centered at the origin.
For $\widehat{\mathsf{A}} (\chi)$, they are arcs
of circumference going from the point $ +i$ to
the point $-i$ through $z$. Finally, for
$\widehat{\mathsf{N}} (\eta)$ the orbits
are circumferences passing through the point
$+ i$ and joining the points $z$ and $-z^\ast$.

For an arbitrary matrix $\mathsf{M}_{as}$ the
orbits can be obtained by transforming the orbits
described before with the appropriate matrix
$\mathsf{C}$ . In Fig.~1.b we have plotted
typical examples of such orbits for elliptic,
hyperbolic, and parabolic actions. We stress
that once the fixed points of the matrix
$\mathsf{M}_{as}$ are known, one can ensure
that $z_a$ will lie in the orbit associated to
$z_s$.

Assume now that we have a finite periodic
structure obtained by repeating $N$ times
the basic period represented by $\mathsf{M}_{as}$.
The overall transfer matrix for this system is
$\mathsf{M}_{as}^N$. In the unit-disk picture,
the transformed field by the $N$-period structure
is represented by the point
\begin{equation}
\label{iterat1}
z_N = \Phi[\mathsf{M}_{as}, z_{N-1}] =
\Phi[\mathsf{M}_{as}^N, z_0] ,
\end{equation}
where $z_0$ is the initial point $z_s$.
The idea that maps iterates could be
applied to this problem was recognized
before, though with a somewhat different
approach~\cite{Fel00,Fel01}.

Henceforth, we shall take $z_0 = 0$,
which is not a serious restriction and
corresponds to the case in which no
light incides from the substrate $[E^{(-)}_s
= 0]$, as it happens usually. Note also
that all the points $z_N$ lie in the
orbit associated to the initial point
$z_0$ by the single period, which is
determined by its fixed points: the
character of these fixed points determine
thus the behavior of the periodic structure.

To illustrate how this geometrical approach
works in practice, we take the single period
as a  Fabry--Perot-like system formed by
two identical plates separated by a spacer
of phase thickness $\delta_2$. This is a
symmetric system for which $R_{as}$ and
$T_{as}$ can be easily computed~\cite{Mon02}.
By varying $\delta_2$ we can choose to work in
the elliptic, the hyperbolic, or the parabolic
case. In Fig.~2 we have plotted the sequence
of successive iterates obtained numerically
for these three regimes.

In the elliptic case, it is clear that
the points $z_N$ revolve in the orbit
centered at the fixed point and the
system never reaches the unit circle.

On the contrary, for the hyperbolic and
parabolic cases the iterates converge to
one of the fixed points on the unit circle,
although with different laws, which correspond
to the band stop and band edges of the system,
respectively~\cite{Lek00}.

\begin{figure}
\centering
\resizebox{0.75\columnwidth}{!}{\includegraphics{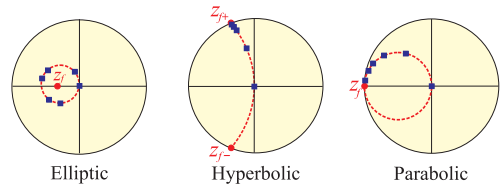}}
\caption{Plot of the successive iterates ($N = 1, \ldots, 5$) for
an elliptic, hyperbolic, and  parabolic action starting from the
origin as the initial point. The physical system is described in
the text. Only hyperbolic and parabolic actions tend to the unit
circle.}
\end{figure}

To gain further insights into these behaviors, we compute
explicitly the $N$th iterate. This can be easily done for the
canonical forms in Eq.~(\ref{Iwasa1}) and then, conjugating as in
(\ref{conjC}) we obtain, after some lengthy calculations, that for
a hyperbolic action one has
\begin{equation}
z_N =  \frac{1- \xi^N }{1 -
\xi^N  (z_{f +}/z_{f -})}z_{f +} ,
\end{equation}
where $ \xi = (\alpha + \beta z_{f -})/(\alpha +
\beta z_{f +})$ is a complex number satisfying
$| \xi | < 1$. Analogously, for the parabolic case
we have
\begin{equation}
z_N = \frac{N  \beta z_f^2}{ N \beta z_f - 1} ,
\end{equation}
where $z_f$ is the (double) fixed point. It
is quite obvious that in both cases $z_N$
converges to one of the fixed points on the
unit circle, so $| z_N | \rightarrow 1$ when
$N$ increases, a typical behavior of perfect
mirror. In the mathematical literature this
limit point is referred to as the Denjoy-Wolff
point of the map~\cite{Kap99}.

To characterize the convergence of $z_N$ we note
that, because $z_0 = z_s= 0$, this initial point
is transformed by the single period into $z_a =
R_{as}$. Therefore, $z_N$ represents the
reflection coefficient of the overall periodic
structure $R_{as}^{(N)}$, which is obviously
different from $(R_{as})^N$. One can then
compute that for the hyperbolic case~\cite{Mon03}
\begin{equation}
|z_N|^2  =
\frac{ | \beta|^2}{|\beta|^2 +
[\sinh ( \chi) / \sinh (N \chi) ]^2 },
\end{equation}
that approaches the unit circle exponentially with
$N$, as one could expect from a band stop, while
for the parabolic action
\begin{equation}
|z_N|^2  =
\frac{ | \beta|^2}{| \beta|^2 + ( 1/ N ) ^2 } ,
\end{equation}
that goes to unity with a typical behavior
$ O(N^{-2})$. This is universal in the physics
of reflection, as put forward in a different
framework by Yeh~\cite{Yeh88} and Lekner~\cite{Lek87}.

To conclude, we expect that the geometrical
scenario presented here could provide an
appropriate tool for analyzing and classifying
the performance of periodic multilayers in an
elegant and concise way that, additionally,
is wider enough to accommodate other periodic
systems appearing in physics.

We acknowledge Jos\'e F. Cari\~{n}ena
and Jos\'e M. Montesinos for illuminating
discussions.


\begin{thebibliography}{99}

\bibitem{dowling}
A complete and up-to-date bibliography
on the subject can be found at the web site
http://home.earthlink.net/\symbol{126}jpdowling/pbgbib.html.

\bibitem{Cen00}
E. Centeno and D. Felbacq,
J. Opt. Soc. Am. A \textbf{17}, 320 (2000).

\bibitem{Fin98}
Y. Fink, J. N. Winn, S. Fan, C. Chen,
J. Michel,  J. D. Joannopoulos, and E. L. Thomas,
Science \textbf{282}, 1679 (1998).

\bibitem{Dow98}
J. P. Dowling,
Science \textbf{282}, 1841 (1998).

\bibitem{Yab98}
E. Yablonovitch, Opt. Lett. \textbf{23}, 1648 (1998).

\bibitem{Chi99}
D. N. Chigrin, A. V. Lavrinenko,
D. A. Yarotsky, and S. V. Gaponenko,
Appl. Phys. A \textbf{68} 25 (1999).

\bibitem{Lek00}
J. Lekner, J. Opt. A  \textbf{2}, 349 (2000).

\bibitem{Gra00}
B. Gralak, G. Tayeb, and S. Enoch,
J. Opt. Soc. Am. A \textbf{17}, 1012 (2000).

\bibitem{Mot00}
M. Notomi, Phys. Rev. B \textbf{62} 10696 (2000).

\bibitem{Yon02}
T. Yonte, J. J. Monz\'{o}n,  L. L. S\'{a}nchez-Soto,
J.~F. Cari\~{n}ena, and C. L\'opez-Lacasta,
J. Opt. Soc. Am. A \textbf{19}, 603 (2002).

\bibitem{Mon02}
J. J. Monz\'{o}n, T. Yonte, L. L. S\'{a}nchez-Soto,
and J.~F. Cari\~{n}ena,
J. Opt. Soc. Am. A \textbf{19}, 985 (2002).


\bibitem{San01}
L. L. S\'{a}nchez-Soto, J. J. Monz\'{o}n,
T. Yonte, and J.~F. Cari\~{n}ena,
Opt. Lett. \textbf{26}, 1400 (2001).


\bibitem{Mon99a}
J. J. Monz\'{o}n and  L. L. S\'{a}nchez-Soto,
Opt. Commun. \textbf{162}, 1 (1999).

\bibitem{Mon99b}
J. J. Monz\'{o}n and  L. L. S\'{a}nchez-Soto,
J. Opt. Soc. Am. A \textbf{16}, 2013 (1999).

\bibitem{Mon01}
J. J. Monz\'{o}n, T. Yonte, and L. L. S\'{a}nchez-Soto,
Opt. Lett. \textbf{26}, 370 (2001).

\bibitem{Fel00}
D. Felbacq,
J. Phys. A \textbf{33} 7137 (2000)

\bibitem{Fel01}
D. Felbacq, B. Guizal, and F. Zolla,
LANL e-print archive Physics/0104074.

\bibitem{Kap99}
J. Kapeluszny, T. Kuczumow, and S. Reich,
Adv. Math. \textbf{143}, 111 (1999).

\bibitem{Mon03}
J. J. Monz\'{o}n, T. Yonte, and  L. L. S\'{a}nchez-Soto,
Opt. Commun. \textbf{218}, 43 (2003).

\bibitem{Yeh88}
P. Yeh, \textit{Optical Waves in Layered Media}
(Wiley, New York, 1988).

\bibitem{Lek87}
J. Lekner, \textit{Theory of Reflection}
(Dordrecht, Amsterdam, 1987).

\end{thebibliography}
\end{document}